# Dark-Exciton-Mediated Fano Resonance from a Single Gold Nanostructure Deposited on Monolayer WS$_2$ at Room Temperature

Mingsong Wang[1], Zilong Wu[2], Alex Krasnok[3], Tianyi Zhang[6], Mingzu Liu[7], He Liu[8], Leonardo Scarabelli[9], Jie Fang[2], Luis M. Liz-Marzán[10,11,12], Mauricio Terrones[6,7,8], Andrea Alù[3,4,5], Yuebing Zheng[*,1,2]

[1]Department of Mechanical Engineering, [2]Materials Science and Engineering Program, Texas Materials Institute, The University of Texas at Austin, Austin, TX 78712, USA
E-mail: zheng@austin.utexas.edu (Y. Z.)
[3]Photonics Initiative, Advanced Science Research Center, [4]Physics Program, Graduate Center, [5]Department of Electrical Engineering, City College of The City University of New York, NY 10031, USA
[6]Department of Materials Science and Engineering, [7]Department of Physics and Center for 2-Dimensional and Layered Materials, [8]Department of Chemistry, The Pennsylvania State University, The Pennsylvania State University, University Park, PA 16802, USA
[9]Department of Chemistry and Biochemistry, California NanoSystems Institute, University of California, Los Angeles, Los Angeles, California 90095, USA
[10]Bionanoplasmonics Laboratory, CIC biomaGUNE, Paseo de Miramón 182, 20014 Donostia-San Sebastián, Spain
[11]Ikerbasque, Basque Foundation for Science, 48013 Bilbao, Spain
[12]Biomedical Research Networking Center in Bioengineering, Biomaterials, and Nanomedicine, CIBER-BBN, 20014 Donostia- San Sebastián, Spain

ABSTRACT: Strong spatial confinement and highly reduced dielectric screening provide monolayer transition metal dichalcogenides (TMDCs) with strong many-body effects, thereby possessing optically forbidden excitonic states (i.e., dark excitons) at room temperature. Herein, we explore the interaction of surface plasmons with dark excitons in hybrid systems consisting of stacked gold nanotriangles (AuNTs) and monolayer WS$_2$. We observe a narrow Fano resonance when the hybrid system is surrounded by water, and we attribute the narrowing of the spectral Fano linewidth to the plasmon-enhanced decay of dark *K-K* excitons. Our results reveal that dark excitons in monolayer WS$_2$ can strongly modify Fano resonances in hybrid plasmon-exciton systems and can be harnessed for novel optical sensors and active nanophotonic devices.

Owing to its significant field enhancement and narrow linewidth, Fano resonances in plasmonic nanostructures have great potential for applications in chemical and biomedical sensing, nonlinear optics and plasmonic nanolasers.[1-3] Many theoretical and experimental works have been carried out to explore Fano resonances that arise from the coupling between surface plasmons (SPs) and excitons in quantum dots (QDs) or dye molecules.[2, 4-9] However, the lack of efficient ways to actively tune the excitonic properties of QDs and dye molecules near plasmonic nanostructures makes the development of active devices based on hybrid plasmon-QD/dye systems challenging. Recently, the interaction between plasmonic nanostructures and emerging two-dimensional (2D) semiconductors, such as monolayer transition metal dichalcogenides (TMDCs), has attracted the attention of various researchers.[10-12] TMDCs possess high carrier mobility, a direct bandgap and strong excitonic and mechanical properties of TMDCs. Combining these outstanding properties with plasmonic nanostructures, able to confine light at the subwavelength scale and generate energetic hot electrons, holds the promise to enhance the performance of TMDC-based optoelectronic components and boost the development of miniaturized and flexible optical devices. For example, plasmonic nanostructures can induce large photoluminescence enhancement of TMDCs via strongly enhanced local electric fields (E-fields) and Purcell phenomena.[13-14]

Electrons and energy can be transferred from plasmonic nanostructures to TMDCs through hot-electron injection and resonance energy transfer, respectively.[15-17] Coupling between exciton states of TMDCs and localized plasmon resonances (LSPRs) can lead to Fano resonance and even strong plasmon-exciton coupling.[18-24] Furthermore, because of the large exciton binding energy and weak dielectric screening, TMDCs have an outstanding tunability regarding their excitonic properties. This tunability further endows hybrid plasmon-TMDC systems with an ability to control Fano interference between surface plasmons and excitons. For example, the tuning of Fano resonance in hybrid plasmon-$WS_2$ systems has been demonstrated through varying the static dielectric constant of the surrounding media.[18]

The strong spatial confinement and low dielectric screening of TMDCs also endow them with exotic excitonic properties, when compared to conventional QDs and dye molecules. For example, density functional theory (DFT) calculations involving many-body perturbation and the Betha-Salpeter equation unravel significant many-body effects in TMDCs.[25] The many-body effects together with the strong spin-orbit coupling in TMDCs lead to the formation of optically forbidden excitonic states (i.e., dark excitons).[26-28] Dark excitons in TMDCs are interesting due to their long lifetimes and strong coupling with bright excitons, thus having great potential in Bose-Einstein condensation, optoelectronic devices and sensing.[26, 29-31] For this reason, TMDCs constitute a promising platform to explore how dark excitons influence plasmon-exciton coupling. In this paper, we demonstrate a tunable narrow Fano resonance stemming from a hybrid plasmon-exciton system by exploiting dark excitons in monolayer $WS_2$.

**Results and Discussion**

Chemically synthesized gold nanotriangles (AuNTs) with an edge length of ~55 nm were selected because they feature atomically flat surfaces, which ensure a strong interaction between AuNTs and TMDCs (a TEM image of AuNTs is shown in Figure 1, inset). AuNTs were synthesized following a previously reported seed-growth method.[32] It should be noticed that there is a layer of cetyltrimethylammonium chloride (CTAC) molecules covering the AuNTs, which prevents direct electron transfer between AuNTs and TMDCs. Monolayer 1H-$WS_2$ flakes were grown on $SiO_2$/Si substrates via chemical vapor deposition (CVD) and were then transferred onto glass slides (detailed information of monolayer $WS_2$ synthesis is provided in the Methods).

Representative optical and AFM images of monolayer $WS_2$ are shown in Figure S1. From the recorded height profiles, we determined a thickness of 0.9 nm, corresponding to a monolayer $WS_2$ flake on a glass substrate.[33-36] An inverted optical microscope equipped with a spectrometer, as schematically illustrated in Figure 1a, was employed to acquire single-nanoparticle scattering spectra (details of optical measurements are provided in the Methods). AuNTs were drop casted on the surface of monolayer $WS_2$.

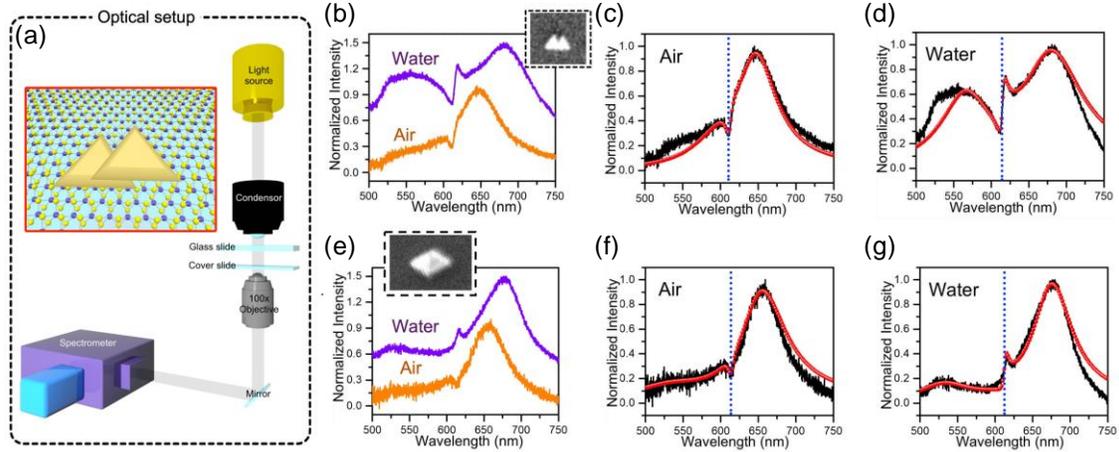

Figure 1. (a) Schematic diagram of the optical setup. Insert is the schematic illustration of stacked AuNTs on monolayer $WS_2$. (b) Scattering spectra of the stacked AuNTs 1 on monolayer $WS_2$ in the air (orange curve) and water (purple curve). Shown in the inset is a SEM image of the stacked AuNTs. Fano fitting of scattering spectra of the stacked AuNTs 1 on monolayer $WS_2$ in the air (c) and water (d). (e) Scattering spectra of the stacked AuNTs 2 on monolayer $WS_2$ in the air (orange curve) and water (purple curve). Shown in the inset is a SEM image of the stacked AuNTs 2. Fano fitting of scattering spectra of the stacked AuNTs 2 on monolayer $WS_2$ in the air (f) and water (g). The blue doted lines in (c), (d), (f) and (g) indicates the peak position of A excitons.

The inserted image in Figure 1a schematically shows the hybrid system of stacked AuNTs on monolayer $WS_2$. The stacked configuration is formed during the drop casting process with one edge of a 55 nm AuNT partially overlapping the one of the other AuNT (SEM images of two stacked AuNTs (stacked AuNTs 1 and 2) on monolayer $WS_2$ are inserted in Figures 1b and e). Figure 1b (e) indicates that scattering spectra of stacked AuNTs 1 (2) in air and water have a peak at ~650 nm (~656 nm) and ~681 nm (~677 nm). To analyze the origin of these peaks, we simulated the scattering spectra of stacked AuNTs on a glass substrate through a finite-difference time domain (FDTD) method. The FDTD simulation results in the Supporting Information indicate that the scattering peaks are associated with the excitation of plasmonic dipole modes.

Scattering spectra in Figures 1b and e also illustrate a clear evolution from a symmetric scattering dip to an asymmetric Fano lineshape at the resonance peak of the A exciton (~615 nm), when the surrounding medium is changed from air to water. This change can be explained by the increased dipole moment of excitons in monolayer $WS_2$ when surrounded by water.[18] When compared to an individual AuNT on monolayer $WS_2$, the stacked AuNTs possess a narrower asymmetric Fano resonance.[18] The maximum value of the asymmetric Fano resonance is at $E_{max}=E_F+\Gamma/(2q)$, the minimum is located at $E_{min}=E_F-\Gamma q/2$, and the width of the resonance is proportional to $E_{max}-E_{min}$, where $E_F$ is the resonant energy, $\Gamma$ is the width of the discrete state, and $q$ is the Fano asymmetry parameter. This suggests that the narrower Fano lineshape reveals a smaller resonance linewidth of A excitons coupled to the stacked AuNTs than the one of regular A excitons.[2]

To interpret the observed spectral change, we fitted the scattering spectra with a Fano approach developed by Gallinet and Martin:[18, 37]

$$\sigma_{scat}(\omega) = \sigma_{ex}(\omega)\sigma_{pl}(\omega) \quad (1)$$

where $\omega$ is the frequency and $\sigma_{scat}(\omega)$ is the scattering cross-section. $\sigma_{ex}(\omega)$ and $\sigma_{pl}(\omega)$ are the excitonic (dark) mode and plasmonic (bright) mode, respectively, which are given by

$$\sigma_{ex}(\omega) = \frac{\left(\frac{\omega^2 - \omega_{ex}^2}{\Gamma_{ex}\omega_{ex}} + q\right)^2 + b}{\left(\frac{\omega^2 - \omega_{ex}^2}{\Gamma_{ex}\omega_{ex}}\right)^2 + 1}, \sigma_{pl}(\omega) = \frac{a^2}{\left(\frac{\omega^2 - \omega_{pl}^2}{\Gamma_{pl}\omega_{pl}}\right)^2 + 1} \quad (2)$$

where $\omega_{ex}$, $\Gamma_{ex}$ and $\omega_{pl}$, $\Gamma_{pl}$ are the resonant frequency and the width of resonance line at half maximum of the exciton and plasmonic resonances, respectively, $b$ is the damping parameter originating from intrinsic losses, $a$ is the maximal amplitude of the resonance. The fitting curves are shown in Figures 1c, d, e and f, and the fitting parameters are listed in Table 1. The results reveal that for the stacked AuNTs 1 (2) |$q$| increases from 0.25 to 0.93 (from 0.00 to 0.70) when the sample was covered by water. It is known that $q=\cot(\delta)$, where $\delta$ is the phase of the time-dependent response function of the excitons, thus playing the role of the continuum at the SP resonance.[38] The corresponding phases obtained from the above Fano fitting for the sample in air and water are $0.58\pi$ and $0.74\pi$ ($0.50\pi$ and $0.69\pi$), respectively. This result is consistent with an increase in the $q$ of a single AuNT on monolayer $WS_2$, when the static dielectric constant increases, reported in our previous study.[18]

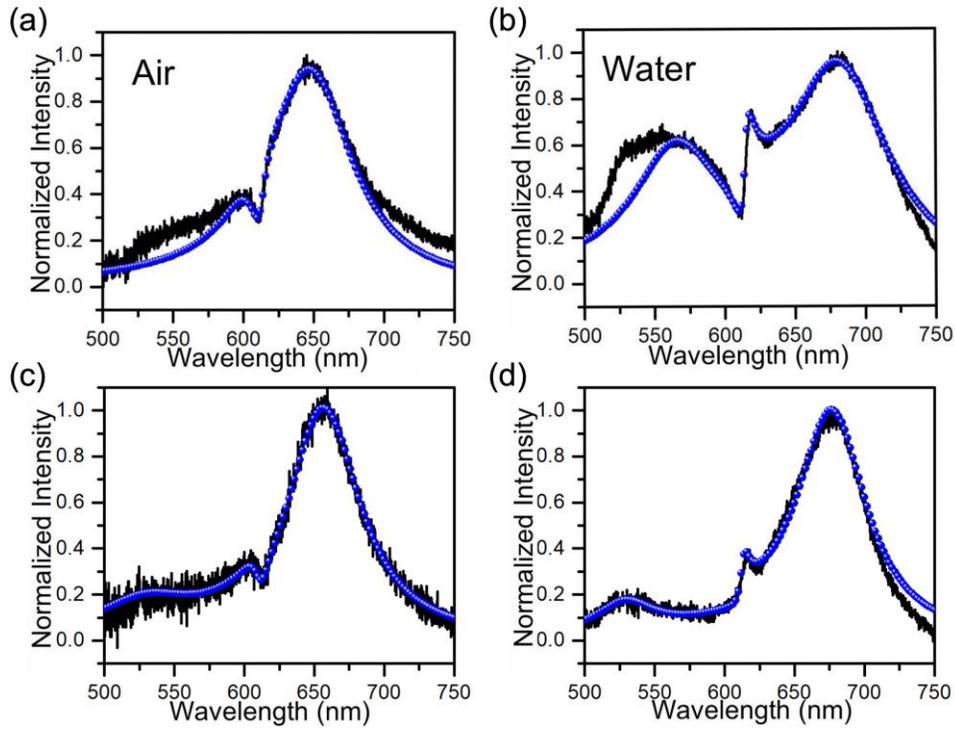

Figure 2. CMT fitting of scattering spectra of stacked AuNTs 1 and 2 on monolayer $WS_2$ in the air (a and c) and water (b and d).

The resonance linewidth values of A excitons listed in Table 1 (31.4 meV and 29.8 meV in the air and 20.8 meV and 24.8 meV in water) are smaller than the reported value (~60 meV).[18, 39-40] To further analyze the experimental result and confirm the reduced resonance linewidth, we applied a modified coupled harmonic oscillator model theory (CMT) to fit our experimental results. CMT has been used to describe the plasmonic analogue of electromagnetically induced transparency (EIT), as well as strong coupling in plasmonic nanostructures.[41] However, in the conventional CMT approach, it is assumed that SPs drive the excitons and the coupled oscillators have the same phase. Thus, the conventional CMT approach cannot be directly used to analyze asymmetric Fano profiles. A phase difference between excitons and SPs should be considered (detailed discussion in Supporting Information), and the corresponding scattering cross-section obtained from the modified CMT is

$$\sigma_{scat} \propto \left| \frac{\omega^2 \tilde{\omega}_{ex}^2}{\tilde{\omega}_{ex}^2 \tilde{\omega}_{pl}^2 - \omega^2 \tilde{g}^2 e^{-2i\varphi}} \right|^2 \tag{3}$$

where $e$ is the elementary charge, $\tilde{g}$ is the plasmon-exciton coupling strength, and $\varphi$ is the phase difference between SPs and the total moment of excitons, $\tilde{\omega}_{ex}^2 = (\omega^2 - \omega_{ex}^2 + i\Gamma_{ex}\omega)$, $\tilde{\omega}_{pl}^2 = (\omega^2 - \omega_{pl}^2 + i\Gamma_{pl}\omega)$. The results obtained by this fitting are plotted in Figures 2a-d, matching the experimental results. The calculated values of the resonance linewidth of A excitons are 28.1 meV and 33.1 meV in air and 22.3 meV and 33.9 meV in water, which are in good agreement with those obtained from the Fano fitting. It is also found that $\tilde{g}$ for the stacked AuNTs 1 (2) is increased from 74.4 meV (53.8 meV) to 95.1 meV (68.2 meV) when the surrounding medium changes from air to water. However, $2\tilde{g}$ for both air and water are smaller than the total loss of the whole system (234.9 meV (223.3 meV) for air and 253.9 meV (191.1 meV) for water), which means that the coupling between the stacked AuNTs 1 (2) and monolayer WS$_2$ does not reach the strong coupling regime when covered by either air or water.

**Table 1**

| Surrounding medium | $q$ | $\delta$ | Linewidth (Fano) | Linewidth (CMT) |
|---|---|---|---|---|
| *Stacked AuNTs 1* | | | | |
| Air | -0.25 | $0.58\pi$ | 31.4 meV | 28.1 meV |
| Water | -0.93 | $0.74\pi$ | 20.8 meV | 22.3 meV |
| *Stacked AuNTs 2* | | | | |
| Air | -0.00 | $0.50\pi$ | 29.8 meV | 33.1 meV |
| Water | -0.70 | $0.69\pi$ | 24.8 meV | 33.9 meV |

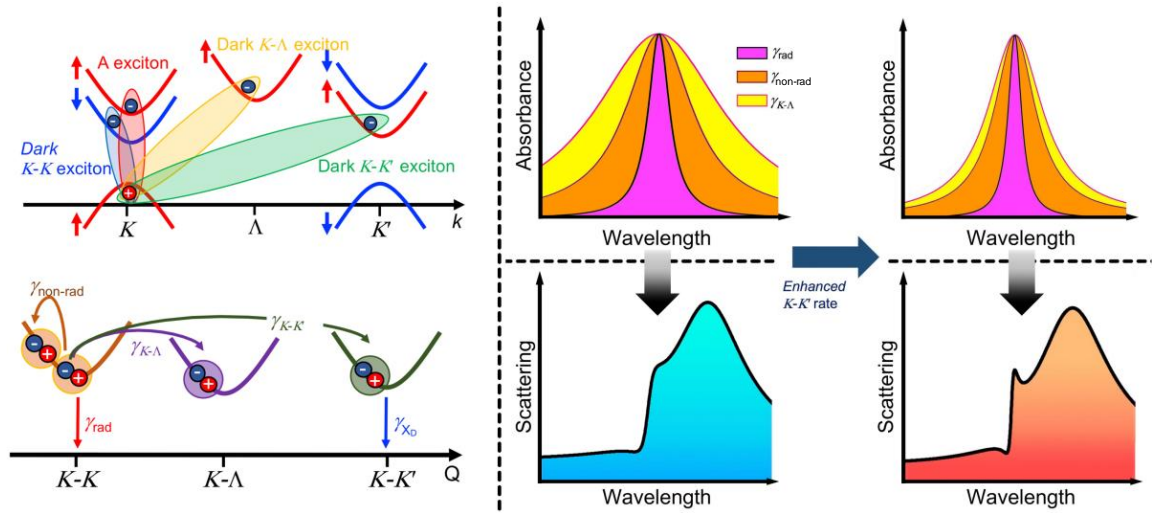

Figure 3. (a) Schematic electronic dispersions at the $K$, $K'$ and $\Lambda$ point. The bright A exciton, intravalley dark $K$-$K$ exciton, intervalley dark $K$-$K'$ and $K$-$\Lambda$ exciton are enclosed by a red, blue, green and yellow oval, respectively. $k$ is the wave vector and red and blue arrows show the direction of the electronic spin. (b) Schematic diagram of the minima of the excitonic center-of-mass motion (**Q**) dispersion $E(\mathbf{Q})$. The arrows illustrate the decay channels of A excitons: the radiative decay ($\gamma_{\text{rad}}$) (red), non-radiative decay ($\gamma_{\text{non-rad}}$) (brown), the A excitons to $K$-$\Lambda$ states ($\gamma_{K-\Lambda}$) (purple) transfer and the A excitons to $K$-$K'$ states ($\gamma_{K-K'}$) transfer (green). The blue arrow represents the decay of the dark $K$-$K$ excitons. ($g_{X_D}$). (c) Schematic diagram showing that the enhanced $\gamma_{K-K'}$ leads to a reduced spectral linewidth for monolayer $WS_2$. (d) Schematic diagram showing that the reduced spectral linewidth of monolayer $WS_2$ leads to a narrower Fano profile in the scattering spectrum.

Recently, Selig et al. have theoretically and experimentally reported that the resonance linewidth of A excitons can be controlled by decay channels from A excitons to intervalley dark excitons.[30] Dark excitons of monolayer WS$_2$ include intravalley dark $K$-$K$ excitons (spin-forbidden excitons) and intervalley dark $K$-$K'$ and $K$-$\Lambda$ excitons (momentum-forbidden excitons), as shown in Figure 3a.[26, 30-31, 42-44] It should be noticed that because at room temperature the photon scattering is faster than the spin flip process, the spin-forbidden transfer from A excitons to dark $K$-$K$ states is negligible in this study.[26, 30] The resonance linewidth of A excitons in monolayer WS$_2$ is mainly affected by the radiative and non-radiative decay rate of A excitons ($\gamma_{rad}$ and $\gamma_{non-rad}$) and A excitons to $K$-$\Lambda$ states rate ($\gamma_{K-\Lambda}$) (all the decay channels are schematically illustrated in Figure 3b), while the A excitons to $K$-$K'$ states rate ($\gamma_{K-K'}$) has a minor contribution to the spectral linewidth owing to the weak electron-phonon coupling element.[30]

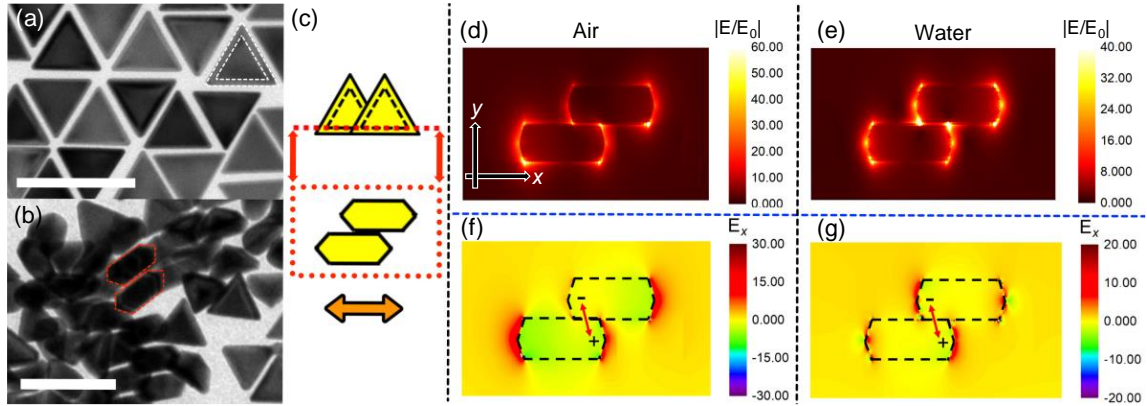

Figure 4. TEM images of AuNTs: (a) The top view of AuNTs shows that there is a noticeable thickness change at the edge of AuNTs, which occurs at the region that is marked by the white dashed triangles; (b) The red dashed polygons illustrate the contour profile of AuNTs, which unravel that the edge of AuNTs is not flat. Scale bars are 100 nm. (c) Schematic images show where E-field is obtained (the red dotted line). The orange arrow shows the polarization of the incident light relative to the stacked AuNTs. (d, e) Simulated E-field distribution of the stacked AuNTs in the air (d) and water (e), respectively. The wavelength of the incident light is 630 nm. (f, g) Simulated distribution of $x$ components of E-field in the air (f) and water (g), respectively. The wavelength of the incident light is 630 nm. Positive and negative signs represent the distribution of positive and negative charges. The red arrow indicates the orientation of the dipole.

In addition, dark $K$-$K'$ excitons degenerate energetically with dark $K$-$K$ excitons.[31] Therefore, when the decay rate of dark $K$-$K$ excitons $X_D$-$g_{X_D}$ increases, the degeneration of dark $K$-$K'$ excitons is enhanced. Park et al. have demonstrated out-of-plane LSPRs profoundly enhance the spontaneous decay rate of dark $K$-$K$ excitons with a large Purcell factor of $\geq 2\times 10^3$, while leaving the decay rates of in-plane A excitons and dark $K$-$\Lambda$ excitons unenhanced.[45] This result implies that the spectral linewidth of A excitons can be controlled by inducing the coupling between out-of-plane LSPRs and dark $K$-$K$ excitons. The plasmon-enhanced decay of dark $K$-$K$ excitons boosts the A excitons to $K$-$K'$ states transfer and leads less A excitons decay through the $K$-$K$ and $K$-$\Lambda$ decay channels, resulting in the resonance linewidth of A excitons ascribed to the $K$-$K$ and $K$-$\Lambda$ decay channels to reduce, as schematically shown in Figure 3c. Figure 3d further illustrates that the reduced resonance linewidth of A excitons can bring a narrower Fano profile in the scattering spectrum of the hybrid system of a single plasmonic nanoparticle on $WS_2$. Thus, we conclude that the decrease in spectral linewidth may be brought by the coupling between LSPRs on the stacked AuNTs and dark $K$-$K$ excitons. In order to verify this assumption, we simulated the scattering spectra and electromagnetic field distribution of the stacked AuNTs using the finite-difference time-domain (FDTD) method (detailed information is in Supporting Information). Herein, we use the stacked AuNTs 1 as an example and the detailed analysis of the stacked AuNTs 2 is in Supporting Information. It should be noticed that the edges of AuNTs are not flat, as shown in Figures 4a and b. The E-field distributions and their $x$ components in Figures 4d-g manifest that under parallel polarized light, the stacked AuNTs 1 possesses out-of-plane dipoles at the resonance peak (~630 nm) of dark $K$-$K$ excitons (about 47 meV below A excitons) in both air and water.[44] While Figures S9 and S10 show the stacked AuNTs 2 possesses out-of-plane dipoles at 630 nm under both parallel and vertically polarized light. Thus, the vertical dipole of the stacked AuNTs 1 and 2 can effectively couple with the out-of-plane dark $K$-$K$ excitons and enhance $g_{X_D}$. The enhanced $g_{X_D}$ boosts $\gamma_{K\text{-}K'}$, which reduces the spectral linewidth.

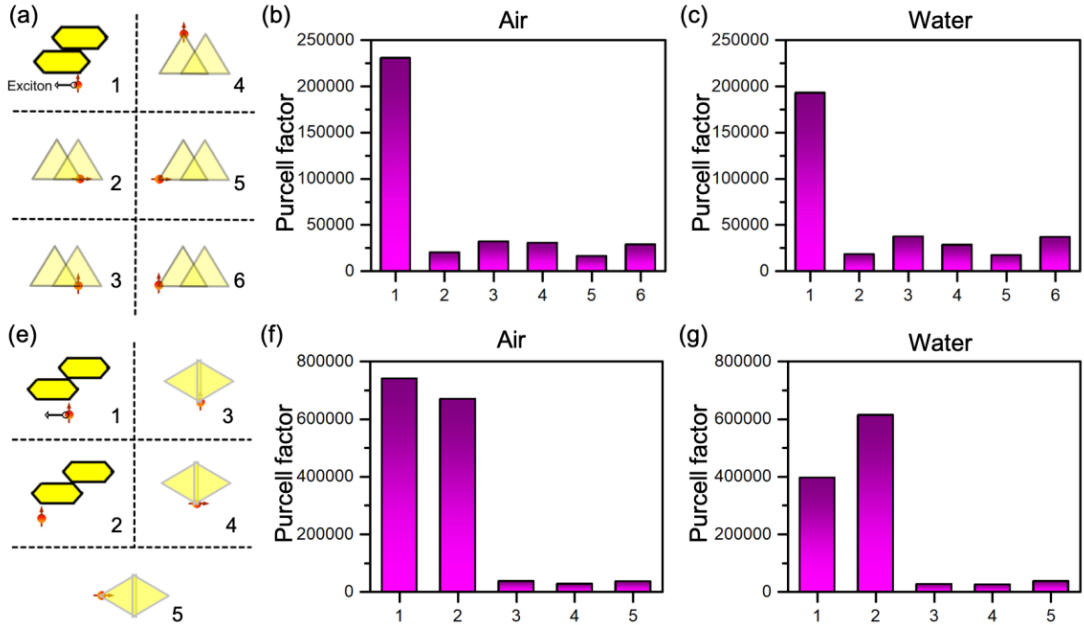

Figure 5. (a) Schematic illustration of six different scenarios in the Purcell factor simulation for the stacked AuNTs 1. The dipole emitters are placed at 1 nm below the stacked AuNTs. Histogram of the simulated Purcell factors in the air (b) and water (c). (e) Schematic illustration of five different scenarios in the Purcell factor simulation for the stacked AuNTs 2. The dipole emitters are placed at 1 nm below the stacked AuNTs. Histogram of the simulated Purcell factors in the air (f) and water (g).

The in-plane SPs of the stacked AuNTs 1 and 2 can also enhance the radiative and non-radiative decay rates of A excitons, thus leading to an increase in spectral linewidth. Therefore, in order to determine whether the coupling between stacked AuNTs and monolayer WS$_2$ leads to a reduced or increased resonance linewidth of A excitons, we compare the plasmon-induced enhancements in the dark $K$-$K$ exciton decay rate ($g_{X_D}$) with the one of the A exciton decay rate ($\gamma_{\text{non-rad}}$ and $\gamma_{\text{rad}}$). The Fermi's golden rule ensures that the spontaneous decay rate of emitters and total Purcell factors (radiative+non-radiative) are directly proportional to the local density states (LDOS).[46] Hence, the simulated Purcell factor can provide a comparison between the plasmon-induced enhancements in the decay rate. Six different scenarios for the stacked AuNTs 1 and five different scenarios for the stacked AuNTs 2 were considered, as schematically shown in Figure 5a and e. The choice of scenarios was based on the locations of the maximum field enhancement at 630 nm for the dark $K$-$K$ exciton and 615 nm for the A exciton, as shown by the E-field distributions depicted in Figure 4 and Figures S4 and S8-10. For the stacked AuNTs 1 (2), the first one (two) corresponds to the enhancement of $g_{X_D}$, whereas others correspond to the enhancement in the sum of $\gamma_{\text{non-rad}}$ and $\gamma_{\text{rad}}$. The obtained Purcell factors for the stacked AuNTs 1 in the air (water) are $2.31\times10^5$ ($1.93\times10^5$), $2.05\times10^4$ ($1.84\times10^4$), $3.21\times10^4$ ($3.75\times10^4$), $3.07\times10^4$ ($2.85\times10^4$), $1.64\times10^4$ ($1.74\times10^4$) and $2.91\times10^4$ ($3.72\times10^4$), respectively (Figures 5b and c). While the obtained Purcell factors for the stacked AuNTs 2 in the air (water) are $7.41\times10^5$ ($3.97\times10^5$), $6.71\times10^5$ ($6.14\times10^5$), $3.76\times10^4$ ($2.71\times10^4$), $2.80\times10^4$ ($2.63\times10^4$) and $3.67\times10^4$ ($3.79\times10^4$), respectively (Figures 5 f and g). This indicates that, upon the excitation of LSPRs, $g_{X_D}$ reaches a larger enhancement than the sum of $\gamma_{\text{non-rad}}$ and $\gamma_{\text{rad}}$. Therefore, the resonance linewidth of A excitons gets reduced, leading to a narrow asymmetric Fano resonance in the scattering spectrum.

**Conclusion**

In summary, we observed a tunable transition from a symmetric dip at the resonance peak of the A exciton to a narrow asymmetric Fano lineshape in the scattering spectra of stacked AuNTs on monolayer $WS_2$ is observed. The Fano fitting and CMT fitting unravel that in the hybrid system, which has out-of-plane LSPRs coupling with dark $K$-$K$ excitons, A excitons have a reduced spectral linewidth when compared to regular A excitons. The FDTD simulation and theoretical analysis reveal this result is derived from the plasmon-enhanced decay rate of dark $K$-$K$ excitons via their coupling with the out-of-plane plasmonic mode in stacked AuNTs. Since A excitons to $K$-$K'$ states transfer has a negligible contribution to the resonance linewidth of A excitons, the enhancement in $g_{X_D}$ reduces the resonance linewidth of A excitons. This makes the Fano interference between the stacked AuNTs and monolayer $WS_2$ to have a narrower spectral linewidth when compared to a single AuNT on monolayer $WS_2$. Our result shows that the coupling between LSPRs and dark excitons can significantly modify the optical properties of TMDCs and paves the way for harnessing the plasmon-dark-exciton coupling for applications such as the biomedical sensing, optical modulators, and plasmonic nanolasers.

**Methods**

**CVD growth of monolayer $WS_2$.** An atmospheric-pressure chemical vapor deposition (APCVD) approach was employed to synthesize monolayer $WS_2$ on $SiO_2$/Si substrate.[47] 5 mg of $WO_3$ powders was first mixed with ~0.5 mg NaBr and then placed on a piece of the $SiO_2$/Si wafer inside an alumina boat. Another $SiO_2$/Si wafer was placed on the top of the boat facing down and then loaded inside a quartz tube. 400 mg sulfur in an alumina boat was placed on the upstream. The furnace was heated up to 825 °C and held for 15 min during synthesis. Meanwhile, a heating belt was used to heat sulfur powders to 250 °C separately. Argon with a flow rate of 100 sccm used as the carrier gas. The AFM images and height profiles of CVD-grown monolayer $WS_2$ were obtained through an atomic force microscope (AFM) (Park Scientific). Monolayer $WS_2$ was then transferred onto glass slides (VWR) via a PMMA-based transfer method.[18]

**Optical measurements.** AuNTs were drop casted onto the top of monolayer $WS_2$. The scattering spectra of single AuNTs on the bare glass substrate or monolayer $WS_2$ were measured through a dark-field setup consisting of an inverted microscope (Ti-E, Nikon), a spectrograph with an EMCCD camera (Andor), and a halogen white light source (12V, 100 W).

**FDTD simulations.** A commercially available software package (FDTD Solutions, Lumerical) was used to conduct FDTD simulations. The scattered light from the AuNTs was collected in a transmission manner and a plane wave was used as the incident light source. The wavelength-dependent dielectric functions of gold were adapted from Johnson and Christy.[48] The stacked configuration is simulated by placing one 55 nm AuNT on the top of another one with a 25 nm overlap at one edge of both AuNTs. Because it is highly possible that the suspended tips of upper AuNTs actually contact with the substrate, cylinders with the same dielectric constant of the glass substrate are added to those tips to removing additional peaks stemming from the asymmetrical dielectric environment.

ASSOCIATED CONTENT

**Supporting Information**.

Optical and AFM images of monolayer $WS_2$; Simulated scattering spectra of stacked AuNTs; E-field disbtributions of stacked AuNTs; the detail information of modified CMT model.

AUTHOR INFORMATION

**Corresponding Author**

*zheng@austin.utexas.edu. (Y. B. Z.)

**Notes**

The authors declare not competing financial interest.

ACKNOWLEDGMENTS

Y. Z. M.W. and Z.W. acknowledge the financial support of the Office of Naval Research Young Investigator Program (N00014-17-1-2424) and the National Science Foundation (NSF-CBET-1704634 and NSF-CMMI-1761743). A.K and A.A. acknowledge support from the Welch Foundation with grant No. F-1802 and the Air Force Office of Scientific Research with MURI grant No. FA9550-17-1-0002.